\documentclass[aps,prl,prabib,twocolumn,showpacs,nofootinbib]{revtex4}
\usepackage{graphicx} \usepackage{amsmath} \usepackage{amssymb}
\usepackage{amsfonts} \usepackage{bm}

\begin{document}

\newcommand{\be}{\begin{equation}} \newcommand{\ee}{\end{equation}}
\newcommand{\bea}{\begin{eqnarray}}\newcommand{\eea}{\end{eqnarray}}

\title{Lower bound of minimal time evolution in quantum mechanics}

\author{Pulak Ranjan Giri} \email{pulakranjan.giri@saha.ac.in}

\affiliation{Theory Division, Saha Institute of Nuclear Physics, 1/AF
Bidhannagar, Calcutta 700064, India}

\begin{abstract}
 We show that the total time of evolution
 from the initial quantum state  to final quantum state and
 then back to the initial state, i.e., making a round trip along the great
 circle over $S^2$,
 must have a lower bound in quantum mechanics, if  the difference
 between two eigenstates  of the $2\times 2$ Hamiltonian is  kept fixed.
 Even the non-hermitian
 quantum mechanics can not reduce it to arbitrarily small value.
 In fact, we show that whether one
 uses a hermitian Hamiltonian or a non-hermitian, the required minimal total
 time of evolution is same. It is argued that in hermitian quantum mechanics
 the condition for  minimal
 time evolution can
 be  understood as a constraint coming from the orthogonality of the
 polarization vector $\bf P$ of the evolving quantum
 state   $\rho=\frac{1}{2}\left(\bf 1+ \bf{P}\cdot\boldsymbol{\sigma}\right)$
 with the vector $\boldsymbol{\mathcal O}(\Theta)$ of
 the $2\times 2$ hermitian Hamiltonians
 $H =\frac{1}{2}\left({\mathcal O}_0\boldsymbol{1}+
 \boldsymbol{\mathcal O}(\Theta)\cdot\boldsymbol{\sigma}\right)$ and it is
 shown that the Hamiltonian $H$ can be parameterized  by two independent
 parameters
 ${\mathcal O}_0$ and $\Theta$.
\end{abstract}



\date{\today}

\maketitle
\noindent
{\bf 1. Introduction}\\

Quantum system is governed by a Hamiltonian $H$ and quantum states
 (we are considering only pure states here) of
the system, belonging to a Hilbert space.  The Hamiltonian acts on
the states of this Hilbert space. In Schr\"{o}dinger picture, the
state evolves and it evolves  in such a way that the norm of the
state remains fixed, i.e., the evolution is unitary. Unitary
evolution  is known to be dictated  by the unitary operator
$\mathcal U= \exp(-itH)$ \cite{reed,dun,landau}. here $t$ is the
evolution time of the system from an initial state to a final state
of the system. In reality, in some situations, this evolution time
has prime importance to think about. For example, in quantum
computation it is desirable to minimize time of evolution of the
orthogonal states of q-bits and it essentially depends on the
transformation speed. The least time to transform one state to other
orthogonal state is known to be $\Delta T= \pi/2E$ \cite{nor,joa},
where $E$ is the energy of the quantum system.

Research work in minimal time evolution has got lots of interest in
recent years \cite{nor,peter,carlini,paulo,martin,brody}, due to its
applicability in quantum  computation
\cite{khaneja1,khaneja2,tanimura, zanardi,pachos,boscain}. Quantum
computation in least possible time is always desirable and  can be
achieved by using time optimal evolution of the quantum states.
Although one can minimize time by designing the gates in a specific
way, time optimal evolution is being thought of as an alternative.

The evolution of a system between two given states in minimum time
is known to be quantum brachistochrone problem \cite{carlini}, a
concept which has come from the brachistochrone problem in classical
mechanics \cite{goldstein}. In hermitian quantum mechanics the
minimal time to evolve a state to other state has a lower bound for
a suitably chosen Hamiltonian provided  the difference between two
eigenstates $E_1, E_2$ ($E_2>E_1$) of the Hamiltonian is  kept
fixed, i.e., $E_2-E_1=2\Delta E= constant$.

The search for  further  minimization of minimal  time evolution
naturally throws research work to the non-hermitian
\cite{bender1,kaushal,cannata,bender2,bender3,bender4,pratick,sinha,
dorey1,dorey2,bender5,ali1,ali2,ali3,ali4,ali5,ali6} 
quantum mechanical domain. The breakthrough
is the recent remarkable work by Bender et al. They showed that with
the same  energy constraint $2\Delta E= constant$, the evolution
time $\Delta T$ of a spin up state to the spin down state under a
particularly chosen non-hermitian $\mathcal{PT}$-symmetric
Hamiltonian $H_{\mathcal{PT}}$, can be made arbitrarily  small.  The
reason behind this peculiar behavior can be understood from the fact
that the states which are orthogonal in hermitian quantum mechanics
under ordinary inner product, are non-orthogonal in non-hermitian
quantum mechanics under $\mathcal{CPT}$ inner product. In fact using
this  $\mathcal{CPT}$ inner product, the distance between the two
states can be made zero in non-hermitian quantum mechanics, although
it has a finite distance in hermitian quantum mechanics. It can be
noted that the shortest evolution time $\Delta T$, the distance
between two states $\Delta S$ and the energy difference $2\Delta E$
between the eigenvalues of the Hamiltonian, which can be considered
as the speed of evolution of the system, can be related by $\Delta
S= 2\Delta E\times\Delta T$ \cite{brody1}. Because of the linear
relation between the distance and evolution time, when the speed is
fixed, one can achieve the evolution in arbitrarily short time if
the distance can be made arbitrarily small.

This arbitrary short evolution time seems to be the result of
introduction of the non-hermiticity and $\mathcal {PT}$-symmetry of
the system. But if one sacrifices  $\mathcal {PT}$-symmetry, is it
still possible to get faster evolution than hermitian quantum
theory? Although without $\mathcal {PT}$-symmetry, the non-hermitian
system generates complex energy eigenvalues, it is however still
possible to get arbitrarily short evolution \cite{paulo}, while
keeping same the energy constraint, $2\Delta E= constant$.

Motivation for our present work comes from the previous works
\cite{carlini,bender,carlini1} on quantum brachistochrone problem in
both hermitian and non-hermitian quantum mechanics. In this article
we ask the most natural question regarding the optimal time
evolution of quantum states. The question is; what is the minimal
time $\Delta T$ to transform one state to another and then transform
back to the same state? Answer to this question is so far not know
for both hermitian and non-hermitian quantum mechanics. In this
article we discuss this issue for both the cases, the hermitian and
the non-hermitian quantum mechanics. We show that the minimal time
required to transform one state to other and then transform back to
the same initial state is  the same, both for hermitian  and
non-hermitian quantum mechanics.

This article is organized as follows: In section 2, we calculate
minimal time to transform one state to its orthogonal state  and
then transform back to the same state in hermitian quantum
mechanics. As a specific case, we consider the simple example of two
state system (it can be a q-bit system or spin half system or other
two state system), which has been discussed  in many places
\cite{penrose,joa,merz, bender}. In section 3, we repeat the same
calculation in non-hermitian $\mathcal P\mathcal T$-symmetric
quantum mechanical domain. We found that in both cases the  minimal
time $\Delta T$ of evolution  is same. We conclude in section 4.

\vskip 0.5 cm
\noindent
{\bf 2. Hermitian quantum mechanics (HQM)}\\

In hermitian quantum mechanics, the generator of time evolution of a
quantum system, which is the Hamiltonian itself, has to be hermitian
$H= H^\dagger$ in order to make the evolution operator $\mathcal U=
\exp(-itH)$ unitary with respect to the ordinary inner product in
Hilbert space. We consider a two state system (it may be a q-bit, or
a spin half system or other two state system), which can be
described by a $2\times 2$ hermitian matrix Hamiltonian. The Hilbert
space of a two state system is the so called Bloch sphere
\cite{penrose}. Two dimensional projective Hilbert space, which is
the set of  one dimensional subspaces of complex two dimensional
Hilbert space is isomorphic to the boundary of Bloch sphere $S^2$,
which is the space of pure states of a two state system. Each pure
state is identified with a point on this sphere $S^2$. Orthogonal
states are identified with the antipodal points on $S^2$. So the
general evolution of a state to another state (not necessarily
orthogonal state) is identified as a curve on $S^2$. The geodesic,
which is the shortest possible paths on $S^2$ between two given
points, is given by the arch length (which should be part of the
great circle through that two points). A generic pure quantum state
vector can be written in terms of the density matrix
$\rho=\frac{1}{2}\left(\bf 1+ \bf{P}\cdot\boldsymbol{\sigma}\right)$
\cite{merz}, where $\bf {P}$$\equiv(P_1,P_2,P_3)$ are real
parameters, $\boldsymbol{\sigma}\equiv(\sigma_1,\sigma_2,\sigma_3)$
are Pauli matrices and $\bf 1$ is the identity matrix. $\bf{P}$ is
called the polarization vector with norm
$P=\sqrt{\boldsymbol{P}^2}=1$. General properties of these density
matrix is that its trace is unity $Tr\left(\rho\right)=1$ and it is
hermitian $\rho=\rho^\dagger$, which is evident from the expression
of density matrix representation $\rho$.

Consider two orthogonal states (i.e., antipodal points on the
sphere) designated by the two density matrix $\rho$ (alternatively
quantum state vector $|\Psi_I\rangle$) and $\tilde{\rho}$
(alternatively quantum state vector $|\Psi_F\rangle$) respectively.
Our objective is to discuss the transformation of the state  $\rho$
to $\tilde{\rho}$ and then back to $\rho$ in minimal time. Here, the
geodesics are the great circles starting from the polar point $\rho$
and passing through antipodal point $\tilde{\rho}$. It can be noted
that the set of geodesics between two antipodal points $\rho$ and
$\tilde{\rho}$ can be parameterized by a parameter
$\Theta\in\mathbb{R}(\mbox{mod}~2\pi)$ of the unitary group $U(1)=
\exp(i\Theta)$. It is thus expected that corresponding to each
geodesic, characterized by a specific value of the parameter
$\Theta$, there exists a hermitian Hamiltonian $H(\Theta)$, which
will transform the state $\rho$ to $\tilde{\rho}$ and back to
$\rho$. The total distance from the point designated by $\rho$ to
$\tilde{\rho}$ and then coming back to again $\rho$ along the
geodesic is the circumference of the great circle through that two
point, which is ${\Delta S}_{\mbox{HQM}}=2\pi$. The maximum speed of
transformation along the geodesic is ${\Delta
V}_{\mbox{HQM}}=E_2-E_1= 2\Delta E$. We now find the total minimal
time ${\Delta T}_{\mbox{HQM}}$ of transformation for
$\rho\to\tilde{\rho}\to\rho$ to be
\begin{eqnarray}
{\Delta T}_{\mbox{HQM}}= {\Delta S}_{\mbox{HQM}}/{\Delta V}_{\mbox{HQM}}=
\pi/\Delta E
\label{time1}
\end{eqnarray}
We now need to find out the  hermitian Hamiltonians
${H}_{\mbox{HQM}}(\Theta)$, which correspond to the minimal time
evolution (\ref{time1}) along the great circles. A general hermitian
$2\times 2$ matrix Hamiltonian on $S^2$ can be written in terms of
four independent parameters $\mathcal O_0$ and $\boldsymbol{\mathcal
O}\equiv(\mathcal O_1,\mathcal O_2,\mathcal O_3)$ as
$H =\frac{1}{2}\left(\mathcal O_0\boldsymbol{1}+
\boldsymbol{\mathcal O}\cdot\boldsymbol{\sigma}\right)$ \cite{merz},
where  $\bf 1$ and $\boldsymbol{\sigma}$ has been defined already.
In our case all four parameters will not  be independent, because we
need to impose some constraints on the Hamiltonian in order to
achieve the desired Hamiltonian which will transform a given state
to its orthogonal state and then back to the original state in
minimal time. The general transformation of a state
$\rho=\frac{1}{2}\left(\bf 1+
\bf{P}\cdot\boldsymbol{\sigma}\right)$, can be associated with the
rate of change of the polarization vector $\bf P$ as $\frac{d\bf
P}{dt}= \boldsymbol{\mathcal O}\times\bf{P}$ \cite{merz} (here
$\times$ is the vector cross product). Obviously the rate of change
will be maximum when the vector $\boldsymbol{\mathcal O}$ is
perpendicular to the polarization vector $\bf P$. So, we identify
two constraints for our Hamiltonian: 1) $2\Delta E= E_2-E_1=
\mathcal O$, where $\mathcal O$ is the norm of the vector
$\boldsymbol{\mathcal O}$. 2) $\boldsymbol{\mathcal O}$ is
perpendicular to the plane of polarization vector through which it
evolves. We also identify two  arbitrariness involved in the
Hamiltonian: 1) The sum of the two eigenvalues $E_1$ and $E_2$,
i.e., $E_1+E_2= \mathcal O_0$ (this is arbitrary) and 2) The
direction of the vector $\boldsymbol{\mathcal O}$, which is confined
in a plane. So, now $\boldsymbol{\mathcal O}$ can be parameterized
with a single parameter, which we identify with our previously
defined parameter $\Theta$. Then, we can immediately write down the
Hamiltonian $H_{\mbox{HQM}}(\mathcal O_0,\Theta)$, which will
transform the state $\rho$ to its orthogonal state in minimal time
as
\begin{eqnarray}
H_{\mbox{HQM}}(\mathcal O_0,\Theta)=\frac{1}{2}\left(\mathcal
O_0\boldsymbol{1}+ \boldsymbol{\mathcal
O}(\Theta)\cdot\boldsymbol{\sigma}\right) \label{ham1}
\end{eqnarray}
This Hamiltonian has only two independent parameters $\mathcal O_0$
and $\Theta$ as it should be. Each value of $\Theta$ corresponds to
a transformation along particular geodesic and for a fixed geodesic,
characterized by $\Theta=\Theta_0$ the transformation can be
achieved by a 1-parameter $\mathcal O_0\in \mathbb{R}$ family of
Hamiltonians $H_{\mbox{HQM}}(\mathcal O_0,\Theta=\Theta_0)$.

\vskip 0.5 cm
\noindent
{\bf 3. Non-hermitian quantum mechanics (NQM)}\\

As mentioned in the introduction, the search for much faster time
evolution than hermitian quantum theory compels one to think beyond
hermitian quantum domain. The obvious choice is the non-hermitian
$\mathcal{PT}$-symmetric quantum mechanics 
\cite{roy1,roy2,roy3,roy4,roy5}, because it is a
consistent quantum theory and can be thought of as an alternative to
the conventional quantum mechanics. Although one may worry about the
unitarity of time evolution, but it can be shown that under the
newly defined inner product, the non-hermitian
$\mathcal{PT}$-symmetric quantum mechanics respects unitarity. So
one can develop quantum brachistochrone problem in this alternative
quantum mechanical setting. Here, our objective is to calculate the
minimal time evolution ${\Delta T}_{\mbox{NQM}}$ from a state $\rho$
(alternatively quantum state vector $|\Psi_I\rangle$) to its
orthogonal state (according to hermitian quantum mechanics)
$\tilde{\rho}$ (alternatively quantum state vector $|\Psi_F\rangle$)
and then back to the same state $\rho$.

But before going into our objective, let us review why in
non-hermitian $\mathcal{PT}$-symmetric theory, transformation
between two orthonormal states (orthonormal according to hermitian
theory) can be made in arbitrarily  short time \cite{bender}. The
clue is in the nontrivial $\mathcal{CPT}$ inner product, defined for
the non-hermitian $\mathcal{PT}$-symmetric quantum mechanics in
order to make the theory unitary. The crucial difference between the
ordinary inner product, defined over Hilbert space in HQM  and
$\mathcal{CPT}$ inner product defined over  Hilbert space in NQM is
that the orthogonal states in  HQM becomes non-orthogonal in  NQM.
This in effect changes the distance ${\Delta S}_{\mbox{NQM}}=
2\cos^{-1}\left(|\langle\Psi_F|\Psi_I\rangle|\right)$
\cite{anandan,bender} between the state $|\Psi_I\rangle$ and
$|\Psi_F\rangle$. In hermitian quantum mechanics, this distance
${\Delta S}_{\mbox{HQM}}$ between two orthogonal states become
${\Delta S}_{\mbox{HQM}}= \pi$, but in non-hermitian quantum
mechanics, the distance between these same two states become
${\Delta S}_{NQM}= \pi-2|\alpha|$ \cite{bender}, where $\alpha$ is a
real parameter dependent on the Hamiltonian $H_{\mbox{NQM}}$ of the
corresponding non-hermitian quantum theory. One can choose a
Hamiltonian $H_{\mbox{NQM}}$ for NQN in such a way that $|\alpha|\to
\pi/2$, thereby making the distance ${\Delta S}_{\mbox{NQM}}\to 0$.
This nontrivial property of the NQM has been capitalized in Ref.
\cite{bender} to show that the minimal time to transform a spin up
state to a spin down state is ${\Delta T}_{\mbox{NQM}}={\Delta
S}_{\mbox{NQM}}/{\Delta V}_{\mbox{NQM}}=0$, where ${\Delta
V}_{\mbox{NQM}}$, the speed of the transformation is kept 
fixed throughout our analysis, both in hermitian and non-hermitian
quantum mechanics i.e., ${\Delta V}_{\mbox{NQM}}= {\Delta
V}_{\mbox{HQM}}=2\Delta E$.

Now we return to our objective, which is to calculate the total
minimal time to transform a state back to its initial state through
the orthonormal (orthonormal in HQM) state. In order to calculate
that, we need to calculate the distance  from a state to its
orthonormal state (orthonormal in HQM) and then back to the initial
state. This total distance  is ${\Delta S}_{\mbox{NQM}}= 2\pi$,
which is independent of the parameter $\alpha$. This shows that in
both cases, hermitian and non-hermitian, the total distance is same.
In fact this total distance is the circumference of the the unit
radius  circle on $S^2$.  We now easily calculate the minimal total
time of transformation to be
\begin{eqnarray}
{\Delta T}_{\mbox{NQM}}= {\Delta S}_{\mbox{NQM}}/{\Delta
V}_{\mbox{NQM}} = \pi/\Delta E\,. \label{time2}
\end{eqnarray}
It is evident from (\ref{time1}) and (\ref{time2}) that in both
hermitian and non-hermitian quantum mechanics, the total minimal
time to transform a state to its orthonormal state and then back to
the initial state, is same. Here we need to clarify our result with
respect to the result of reference \cite{bender}. It can be easily
understood that our result is in  agreement with
\cite{bender}. Divide the total distance  ${\Delta S}_{\mbox{NQM}}$
for the transformation $\rho\to\tilde{\rho}\to\rho$ into two parts.
One ${\Delta S}_{\mbox{NQM1}}=\pi-2|\alpha|$ for the transformation
$\rho\to\tilde{\rho}$ and other ${\Delta
S}_{\mbox{NQM2}}=\pi+2|\alpha|$ for the transformation
$\tilde{\rho}\to\rho$. If we make ${\Delta S}_{\mbox{NQM1}}=0$ by
tuning $|\alpha|\to\pi/2$, then the other distance becomes ${\Delta
S}_{\mbox{NQM2}}= 2\pi$. The total distance however remains fixed
i.e., ${\Delta S}_{\mbox{NQM}}={\Delta S}_{\mbox{NQM1}}+{\Delta
S}_{\mbox{NQM2}}=2\pi$ and total minimal  time thus has a lower
bound $\pi/\Delta E$.

\vskip 0.5 cm
\noindent
{\bf 4. Conclusion and discussion}\\

 We have shown that to  make a round trip along the great
 circle over $S^2$, the Hamiltonian needs a minimal time which has 
 a lower bound $\Delta T=\pi/\Delta E$ in quantum mechanics. Non-hermitian
 quantum mechanics can not reduce it to arbitrarily small value.
 We have also shown that the total minimal time is same in both quantum
 mechanics. Our result is shown to be in agreement with the
 recent work \cite{bender} on quantum brachistochrone problem.
 We have also shown  that in hermitian quantum mechanics
 the condition for  minimal
 time evolution can
 be considered as a constraint coming from the orthogonality of the
 polarization vector $\bf P$ of the evolving quantum
 state   $\rho=\frac{1}{2}\left(\bf 1+ \bf{P}\cdot\boldsymbol{\sigma}\right)$
 with the vector $\boldsymbol{\mathcal O}(\Theta)$ of
 the $2\times 2$ hermitian Hamiltonians
 $H =\frac{1}{2}\left({\mathcal O}_0\boldsymbol{1}+
 \boldsymbol{\mathcal O}(\Theta)\cdot\boldsymbol{\sigma}\right)$.
 The Hamiltonian $H$ can be parameterized  by two independent
 parameters ${\mathcal O}_0$ and $\Theta$. ${\mathcal O}_0$ has been
 identified as the sum of the eigenvalues of the Hamiltonian and $\Theta$ is
 responsible for evolution along different great circles through the two given
 antipodal points. After we submitted our work to arXiv   others also
 reported in this subject \cite{ali7,guen,ali8,nes,fry,rotter}.

\vskip 0.5 cm

\noindent
{\bf Acknowledgment}\\

Author thanks P. B. Pal for carefully reading the manuscript
and suggesting  imrovements of the manuscript. Author also 
acknowledges the valuable comments on the manuscript made by P. Mitra.

\end{document}